\documentclass[
aps, 
reprint, 
longbibliography
]{revtex4-2}
\usepackage[USenglish]{babel}
\usepackage{tikz,calc,float}
\usepackage{amsmath,amsthm,amsfonts,amssymb,braket}
\usepackage{dsfont}
\usepackage{graphicx}

\usepackage{hyperref}

\newcommand{\christ}[2]{{\Gamma^{#1}}_{#2}}
\DeclareMathOperator{\tr}{tr}

\newcommand{\dd}{{\mathrm{d}}}

\DeclareMathOperator{\I}{I}
\DeclareMathOperator{\II}{II}

\definecolor{color1}{RGB}{0,80,155}	
\definecolor{color2}{RGB}{0,117,223}

\begin{document}
	
	\title{The evolution of expanding spacetime realizes approximate quantum cloning}
	
	\author{Laura Niermann}
	\email{laura.niermann@itp.uni-hannover.de}
	\affiliation{Institut f\"ur Theoretische Physik, Leibniz Universit\"at Hannover, Appelstr. 2, 30167 Hannover, Germany}
	
	\author{Tobias J.\ Osborne}
	\affiliation{Institut f\"ur Theoretische Physik, Leibniz Universit\"at Hannover, Appelstr. 2, 30167 Hannover, Germany}
	
	\begin{abstract}
	We investigate how quantum information, encoded in a quantum field, evolves during the expansion of spacetime. Due to information loss across the horizon, a local observer experiences this evolution as a nonunitary quantum channel. We obtain this channel in the case of de Sitter spacetime by assuming the initial global state encodes a signal state via fluctuations of the Bunch-Davies vacuum. Notably, de Sitter evolution exhibits intriguing cloning properties, establishing a connection between the curvature of spacetime and the propagation of quantum information.
	\end{abstract}
	

	\maketitle

\section{Introduction}

Cosmological observations currently favor the hypothesis that we live in an asymptotically de Sitter (dS) universe \cite{Schmidt1998, Perlmutter1999, Carmeli2001, Peiris2003}, implying that it will approach a state of accelerated expansion for late times. Thus, it is crucial to understand the properties of dS in the context of quantum gravity. This presents significant challenges because the boundaries of dS are located at timelike infinity so to understand the quantum evolution of fields in dS there are still many mysteries to be understood \cite{Witten2001,  Bousso1999a, Bousso2000, Bousso2001, Balasubramanian2001, Balasubramanian2002, Strominger2001, Strominger2001a, Hawking2001}. 

The quantum nature of de Sitter has received growing attention in recent years, with new input from quantum information theory and tensor networks in the context of the AdS/CFT correspondence  \cite{Maldacena1998,Maldacena1999,Susskind1995,Gubser1998,Witten1998,Bousso2002a,Pastawski2015}. Here, toy models \cite{Kunkolienkar2017,Bao2017,Milsted2018,Niermann2022} have provided fertile ground for new ideas leading to the hypothesis that time evolution for expanding spacetimes is isometric and nonunitary \cite{Cotler2022,Cotler2023}. 
The physical justification for this hypothesis arises from the belief that there are no degrees of freedom localized on lengthscales smaller than the Planck length, so the dimension of Hilbert space in an expanding spacetime \emph{increases} with time \footnote{Conservation of probability requires only that the evolution is isometric.}. It is now an intriguing question to study how quantum information is created and propagates in such a setting. 

Here, we investigate the nonunitary evolution of a quantum state as spacetime expands. In particular, we consider the proposal that, as spacetime expands, preexisting quantum information is redundantly ``copied'' into emerging degrees of freedom. While the no-cloning theorem \cite{Wootters1982} forbids the copying of unknown quantum states, it is possible to approximately clone quantum information, and there is extensive literature on such optimal cloning channels \cite{Buzek1996,Gisin1997,Bruss1998,Buzek1998,Werner1998}. In the context of quantum fields in dS, we discover that, indeed, time evolution realizes such an approximate cloning channel. Our analysis draws inspiration from and is strongly informed by extensive investigations of the Unruh effect in Minkowski spacetime, thanks to the strong parallel between the Unruh setting and dS: local observers in dS are unavoidably accelerating away from each other. There is one crucial difference, however: in the case of the Unruh effect, spacetime is flat, and the horizon witnessed by an accelerated observer is artificial, whereas in dS, all local observers agree that there is a horizon.

The nonunitary nature of \emph{local} quantum mechanical evolution in dS is a consequence of the locally thermal character of vacuum states in curved spacetime, which, in turn, directly arises from entanglement between causally disconnected regions \cite{Higuchi2018}. This phenomenon was first studied for flat spacetime in \cite{Higuchi1987, Fulling1973, Davies1976}, and further investigated in many references such as \cite{Crispino2008} and \cite{Daicic1992}. The case of dS has been taken up in numerous studies \cite{Banerjee2019, Casadio2011, Garbrecht2004, Jia2015, Deser1999, Deser1997, Santos2021, Kim2016, Jacobson1998, Jennings2005, Lapedes1978, KamesKing2022, Das2001, Acquaviva2011, Ahmed2023, Markkanen2018, Yu2011, Zych2020}, where the Unruh effect was used to assign thermal properties to dS. 

The Unruh effect in flat spacetime has also been the focus of recent research investigating the fate of quantum information encoded in logical qubits arising from fluctuations of quantum fields \cite{Bradler2009, Bradler2012, Bradler2010}. These works led to the striking discovery that such encoded information, as witnessed by an accelerated observer, is modeled by the output of a quantum \emph{Unruh channel}, interpreted as a block-diagonal sum of optimal cloning channels. This remarkable observation is the basis for the present work; we study the properties of the analogous channel in the dS setting and find that it exhibits similar cloning properties. The core novelty here, however, is that the dS quantum channel enjoys a fundamentally different physical interpretation: The cloning property can be directly associated with the expansion of spacetime. 

This paper is organized as follows: In Sec.~\ref{sec:localObserver}, we characterize a local observer in dS spacetime and introduce the static coordinate system best suited to describe local physics in de Sitter. We then review the quantization of a scalar field in dS in Sec.~\ref{sec:quantizationStatic}, including the Bunch-Davies vacuum state. In Sec.~\ref{sec:trafoMultiRail}, we describe how logical quantum information is encoded via perturbations of the Euclidean vacuum state. We then derive the corresponding Unruh channel in Sec.~\ref{sec:channel}, which describes the evolution of the initial state as perceived by a local observer. Finally, in Sec.~\ref{sec:cloning}, we relate the Unruh channel to optimal cloning channels. The appendices present additional calculational details.

\section{Local observers in de Sitter} \label{sec:localObserver}
\begin{figure}
	\centering
	\includegraphics[page=1,width=\linewidth]{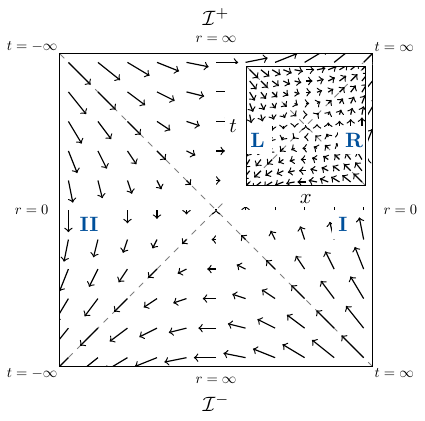}
	\caption{Comparision between Killing fields in de Sitter spacetime (left) in static coordinates and flat spacetime with an accelerated observer (insert):  Killing fields in the Penrose diagram of dS$_3$ where the left and the right boundary are timelike lines associated with the poles in static patch I and II respectively. The static patches are bounded by the dashed line at $r^2=\ell^2$. A horizontal slice of the dS$_3$ Penrose diagram is associated with a sphere, which is de Sitter space at that given time slice, and each point is associated with a circle. The Killing field of a uniformly accelerated observer in Minkowski spacetime take a similar form.}
	\label{fig:Killing-dS-patch}
\end{figure}

There are several different ways to quantize de Sitter spacetime. 
The physics accessible to a local observer can be conveniently described in static coordinates as argued, e.g., in \cite{Witten2023}. In this paper, we work with static coordinates, where $2+1$ dimensional de Sitter spacetime with de Sitter radius $\ell$ is embedded in $3+1$ dimensional Minkowski space:
\begin{align}
	\begin{split}
		z_0 =& \sqrt{\ell^2-r^2} \sinh(t/\ell),\\
		z_1 =& (-1)^\text{patch} \sqrt{\ell^2-r^2} \cosh(t/\ell),\\
		z_2 =& r \cos\theta,\\
		z_3 =& r \sin\theta.
	\end{split}
\end{align}
Here, the physics experienced by a local observer is similar to what we are familiar with from flat spacetime.
A crucial reason for this coordinate choice is that there is no globally timelike Killing field in de Sitter, however, there is a timelike Killing field in a single static patch, as shown in Fig.~\ref{fig:Killing-dS-patch}. The timelike Killing field in a static patch makes it possible to define a notion of time for the local observer living in one static patch, however, this differs between static patch $\I$ and $\II$. While time runs forward in one patch, it runs backward in the other, as emphasised, e.g., in \cite[p. 20]{Witten2023}. 
Time running in different directions in different parts of spacetime does not cause conflict for a local observer because both static patches are causally disconnected.

The causal structure of de Sitter spacetime is similar to that of the Rindler setting, where the static patches in de Sitter are analogous to the Rindler wedges for flat space. A key difference, however, is that in the flat spacetime setting, the Killing trajectories of inertial observers are also geodesics, while for dS this is only the case for the trajectory with $r=0$, which can be seen from the absolute value of the proper acceleration of an observer following a Killing trajectory (see App.~\ref{app:KillingTrajectories}):
\begin{align}
    a_\mu  a^\mu=\frac{1}{\ell^2-r^2} \frac{r^2}{\ell^2}. 
\end{align} The closer the trajectory is to the horizon, the larger its proper acceleration becomes, diverging for $|r|\rightarrow|\ell|$. For simplicity, for the rest of this paper we only consider the case of de Sitter radius $\ell=1$. The generalization to other radii is straightforward.

\section{Quantum fields in dS} \label{sec:quantizationStatic} 

The quantization of scalar fields in dS has been comprehensively described in numerous references. 
Here, we only supply a brief review following Bousso, Maloney, and Strominger \cite{Bousso2001}, as well as Higuchi and Yamamoto \cite{Higuchi2018}.

There is an ambiguity in the choice of vacuum state in curved spacetime \cite{Birrell1982}. We use a quantization of dS where the positive-frequency mode operators $\hat{a}_{k \omega}$, $\omega\ge 0$, annihilate the \emph{Bunch-Davies} vacuum state \cite[Eq.~(88)]{Higuchi2018}:
\begin{align}
    \ket\Omega =& \mathcal{N} \exp\left[\sum_k \int_0^\infty d\omega e^{-\pi\omega}
    (\hat{a}_{k \omega}^{\I})^\dagger (\hat{a}_{k \omega}^{\II})^\dagger \right] \ket{\Omega_{\I}} \otimes \ket{\Omega_{\II}},
\end{align}
where $\hat{a}_{k \omega}^{\I}$ and $\hat{a}_{k \omega}^{\II}$ are the annihilation operators associated with static patches $\I$ and $\II$, respectively, with $\ket{\Omega_{\I}}$ and $\ket{\Omega_{\II}}$ the corresponding local vacuum states.
The choice of the Bunch-Davies vacuum state is reasonable as de Sitter Unruh detectors respond thermally, it is invariant under the full de Sitter group, and it can be expressed as a linear combination of creation operators associated with the static patches $\I$ and $\II$ as derived in \cite[section 6]{Bousso2001}. (The continuation to the lower Euclidean hemisphere associated with the past of de Sitter is analytic. This is why the Bunch-Davies vacuum is also referred to as the Euclidean vacuum state.)

The positive-frequency mode functions associated to the Bunch-Davies vacuum state can be expressed as a linear combination of the mode functions of both static patches as shown in \cite{Bousso2001}:
\begin{align}\begin{split}
	\phi_{k \omega}=& \phi_{k \omega}^{\I} + e^{-\pi\omega} (\phi_{k \omega}^{\II})^\ast \label{eq:modeFktGlobal}.\\ 
\end{split}\end{align}
Note that the role of positive and negative frequencies in static patch $\II$ are flipped with respect to static patch $\I$. The global mode functions in Eq.~(\ref{eq:modeFktGlobal}) are a combination of positive and negative frequency modes from  different static patches. The corresponding annihilation operators annihilating the Bunch-Davies vacuum state are thus 
\begin{align}\begin{split}
	\hat{a}_{k \omega} =& \frac{1}{\sqrt{1-e^{-2\pi\omega}}} \left( \hat{a}_{k \omega}^{\I} - e^{-\pi\omega} (\hat{a}_{k \omega}^{\II})^\dagger \right).
\end{split}\end{align}
A second linearly independent set of positive frequency modes can be defined by interchanging static patches. 

\section{The initial state of the de Sitter universe} \label{sec:trafoMultiRail}
To analyse the propagation of quantum information in dS we encode an initial ``signal'' or ``message'' logical quantum state into perturbations of the Bunch-Davies state. These fluctuations are mediated by globally created particles. Thus the initial state of the universe is given by the following  \emph{multi-rail state}: 
\begin{align}
	\ket{\psi} = \sum_{j=1}^{d} c_j (\hat{a}_{k_j \omega_j})^\dagger \ket{\Omega},
	\label{eq:multiRail}
\end{align}
where $k_i$ are the temporal frequencies for $i=1,\dots,d$, which label the $d$ angular harmonic modes at fixed frequency $\omega$. One interpretation of this state is that it is created by a \emph{superobserver} so that at the bottleneck of dS (occurring at $t=0$) the state is encoded across the entire temporal slice. A similar procedure was described in \cite{Bradler2012} in the case of flat spacetime to encode a logical qubit for use as an initial state of the Unruh channel.

As a consequence of the squeezed-state structure of the Bunch-Davies vacuum we can express the multi-rail state purely in terms of static patch modes (the detailed derivation is provided in App.~\ref{app:multi-rail-expressions}):
\begin{align}
    \ket\psi 
	=& \sum_{j=1}^{d} c_j \sqrt{1-e^{-2\pi\omega_j}} (\hat{a}_{{k_j \omega_j}}^{\I})^\dagger  \ket{\Omega}. \label{eq:multi-rail-E}
\end{align} 
Exploiting the multinomial theorem and multi-index notation, we find 
\begin{align}
	\ket{\psi} 
	=& \sum_{j=1}^{d} c_j \sqrt{1-z_j^2}  \sum_{n=0}^\infty   \left(\prod_{\omega}^{} \sqrt{1-z^2} z^n\right) \nonumber \\ & \quad \cdot \sum_{L_n} \sqrt{l_j+1} \ket{L^{(j)}}_{\I} \ket{L}_{\II},\label{eq:multi-rail-E-new}
\end{align} 
where $z=e^{-\pi\omega}$, $l_j$ are occupation numbers, and $\ket{L^{(j)}}_{\I}$ and $\ket{L}_{\II}$ are certain multimode Fock states (for the precise definitions, and derivation, see App.~\ref{app:multinomial}).

\vspace*{.5cm}
\section{Unruh channel} \label{sec:channel}
In this section we derive the quantum channel (a nonunitary completely positive map) which describes how the initial state $|\psi\rangle$ of the scalar field in dS appears to local observers in a single static patch. On a superficial mathematical level the calculations follow the comparable situation for accelerated observers in Minkowski spacetime: To calculate the Unruh channel, we trace out from $|\psi\rangle$ all the modes from static patch $\II$, which are precisely those which cannot influence the channel for an inertial observer in static patch $\I$:
\begin{align}
	\mathcal{E}(\psi)  =& \tr_{\II}( \ket{\psi}\bra\psi )= \rho_{\I} \otimes \rho_{\I}^\text{residual}\label{eq:cloning-channel}
\end{align}
Here the \emph{residual part} $\rho_{\I}^\text{residual}$ is the state of the subsystem defined by those modes satisfying one of the two following conditions: (i) the temporal frequency $k_i$ is not equal to $\omega$ (ii) the temporal frequency is equal to $\omega$, but $k\neq k_i$. By neglecting the residual part we are left with a subsystem comprised of only a finite number of modes.  Physically, this can be interpreted as the state perceived by a detector tuned to exactly these output modes. This situation is similar to that encountered when studying Rindler quantization \cite{Bradler2009}. The action of the Unruh channel is then obtained according to the following additional assumption, namely, that the frequencies $\omega_k$ are chosen so close that all $z_i=e^{-\pi\omega}$ are approximately equal (to $z$). 
Accordingly, the output density operator $\rho_{\I}$, derived in App.~\ref{app:cloning-channel}, is given by
\begin{align}
        \rho_{\I}=& (1-z^2)^{d+1} \bigoplus_{n=0}^{\infty} z^{2n} \sigma_{\I}^{(n)}, 
    \end{align}
    with \begin{align}
    \sigma_{\I}^{(n)} =& \sum_{j=1}^d |c_j|^2(l_j+1)\sum_{L_n} \ket{L^{(j)}}\bra{L^{(j)}}_{\I} \nonumber\\&+ \sum_{\substack{j,\tilde{j}=1\\j\neq\tilde{j}}}^d c_j c_{\tilde{j}}\sqrt{l_j+1}\sqrt{l_{\tilde{j}}+1}\sum_{L_n} \ket{L^{(j)}}\bra{L^{(\tilde j)}}_{\I}. \label{eq:channel-block-cloning}
\end{align}
It is noteworthy that the output state is realised by the action of a direct sum of channels. As we explain in the next section these each realise $n$-dimensional (approximate) cloning channels.

\section{Cloning properties of the channel} \label{sec:cloning}
The Unruh channel derived in the previous section has a block-diagonal structure, with the output state comprised of the blocks Eq.~(\ref{eq:channel-block-cloning}). As described in \cite{Bradler2010} in the Rindler context, each block density operator can be interpreted as an instance of a $1\rightarrow n$ cloning channel acting on the initial state.

Cloning, in general, refers to the process of creating identical copies of an arbitrary unknown state. In quantum mechanics, it is impossible to create a perfect copy of an unknown state due to the \emph{no-cloning theorem} \cite{Wootters1982}. Although perfect cloning is impossible, it is nevertheless possible to make imperfect copies. This is why we always refer to \emph{optimal cloning} when speaking about cloning in the context of quantum mechanics. 
A \emph{quantum cloning machine} thus acts on an unknown quantum state and performs a transformation to generate (imperfect) copies of the original state. Optimal cloning machines were introduced in \cite{Buzek1996} for arbitrary states from a quantum mechanical spin-$\frac12$ system. These were shown to be optimal in the following years in \cite{Gisin1997, Bruss1998, Werner1998} and further generalized in \cite{Buzek1998} to higher-dimensional systems. In general, a cloning machine is specified by the following conditions, as characterized in \cite{Buzek1998}: 
After the quantum cloning machine has performed the transformation, both the initial state and its copy are in the same state. All pure initial states have to be copied equally well, and even though the copies cannot be perfect, they have to be as close to the initial state as possible to ensure optimal cloning. To realise this, an $N$-dimensional cloning machine is prepared in an initial state $\ket{X}_c$. The cloning transformation is then described by a unitary operator acting on a basis $\ket{\psi_i}_a$ of the original quantum system, an $N$-dimensional quantum system prepared in some fiducial state $\ket0_b$, and the quantum cloner internal state $\ket{X}_c$ as described in \cite[Eq.~(9)]{Buzek1998}):
\begin{align}
	\ket{\psi_i}_a \ket{0}_b \ket{X}_c &\overset{U}{\longrightarrow} \alpha \ket{\psi_i}_a \ket{\psi_i}_b \ket{x_i}_c \nonumber \\&+ \beta \sum_{j\neq i}^N \left(\ket{\psi_i}_a \ket{\psi_j}_b + \ket{\psi_j}_a \ket{\psi_i}_b \right)\ket{x_j}_c
\end{align}where the parameter $\alpha$ and $\beta$ are real parameters that have to fulfill $\alpha^2 +2(N-1)\beta^2=1$ to ensure a unitary cloning transformation. After tracing out the cloner internal subsystem, this cloning transformation then generates the following output state for each subsystem:
\begin{align}
	\hat{\rho}_a^{(\text{out})} =& s \ket\psi\bra\psi_a + \frac{1-s}{M}\mathds{1} \label{eq:cloning-scaled}\\
 	=&\sum_{i=1}^N |\gamma_i|^2 \left(\alpha^2 + (d-2) \beta^2 \right) \ket{\psi_i}\bra{\psi_i} \nonumber\\& + \sum_{\substack{i,j=1\\ i\neq j}}^{N} \gamma_i \gamma_j^\ast \left(2 \alpha\beta + (d-2) \beta^2 \right) \ket{\psi_i}\bra{\psi_j} + \beta^2 \mathds{1}, \label{eq:cloning-n}
\end{align}where the scaling factor $s=\frac{N+2}{2(N+1)}$ characterizes the quality of the clones depending on the dimension. To bring the output state in the scaled form from Eq.~\ref{eq:cloning-scaled}, the parameter has to satisfy further conditions. 

The blocks $\sigma_{\I}^{n}$ in Eq.~(\ref{eq:channel-block-cloning}) have a structure identical to $\hat{\rho}_a^\text{out}$ in Eq.~(\ref{eq:cloning-n}) and can therefore may be interpreted as instances of the action of $n$-dimensional cloning channels. 

There is a compelling physical interpretation of these observations. According to the key assumption referenced in the introduction, namely, that the dimension of Hilbert space in an expanding spacetime \emph{increases} with time, we must determine which states the emerging degrees of freedom find themselves in. There are a couple of naive options, e.g., maybe the new degrees of freedom are initialised in some vacuum-type state, or perhaps in some thermal state. This is natural enough, but it only explains the vacuum situation. What if the state of a quantum field is not in the vacuum (as is presumably the case for our universe)? In this case, adding in new degrees of freedom in some independent product state is unlikely to be compatible with basic symmetries (it would break translation invariance, etc.) However, by filling the new degrees of freedom emerging from the expansion of spacetime with (approximate) \emph{copies} of an initial state $|\psi\rangle$ we are able to preserve symmetries, e.g., translation invariance. 
	
\section{Conclusion}
In this paper, we have investigated how a global initial state, realised via perturbations of the Bunch-Davies vacuum, for a scalar quantum field evolves in a static patch, associated with a local observer. The acceleration intrinsic to de Sitter spacetime results in an Unruh channel, which acts as an approximate cloning channel on the logical input state. 

There are many questions to be explored: One option for future research would be to study more properties of the channel and thus relate the curvature of dS to the channel capacity of the Unruh channel. In particular, does the flat limit somehow reduce to the Rindler case in a natural way? Another would be to use the Unruh channel as a building block for tensor-network models of dS, such as those introduced in \cite{Niermann2022}.

\section*{Acknowledgments} Special thanks go to Jordan Cotler who suggested this research direction, who supplied key derivations, and for various interesting, insightful, and valuable discussions. Without your help, this paper would not have been possible! This work was supported, in part, by the Quantum Valley Lower Saxony (QVLS), the DFG through SFB 1227 (DQ-mat), the RTG 1991, and funded by the Deutsche Forschungsgemeinschaft (DFG, German Research Foundation) under Germany's Excellence Strategy EXC-2123 QuantumFrontiers 390837967.

	\begin{appendix}
		\section*{Appendix}
		

\section{Properties of Killing trajectories}\label{app:KillingTrajectories}
The static metric in $2+1$ dimensional de Sitter spacetime is \begin{align}
    ds^2 = - \left(1-\frac{r^2}{\ell^2}\right)^2 dt^2 + \frac{\ell^2}{\ell^2-r^2} dr^2 + r^2d\theta^2
\end{align}

We consider trajectories in static coordinates with constant $r$ coordinate: \begin{align}
x_r(s) = \begin{pmatrix}
	t(s) \\ r(s) \\ \theta(s)
\end{pmatrix} = \begin{pmatrix}
s\\r_i\\\theta_0
\end{pmatrix}
	\end{align}
where $r_i$ is constant and the choice of $r_i$ determines the trajectory. The tangent vector (since the only change happens in the time coordinate) is $t^\mu = \begin{pmatrix} 1\\0\\0 \end{pmatrix}$ which gives us the proper time 
\begin{align}
	\tau =& \int \sqrt{-g_{ab} t^a t^b} \dd t = \int \sqrt{-g_{tt}}\dd t \nonumber \\
	=& \int \sqrt{1-\frac{r_i^2}{\ell^2}}\dd t = \sqrt{1-\frac{r_i^2}{\ell^2}} t
\end{align} where $t$ is the coordinate time in static coordinates. It needs to be noted that the proper time differs for different trajectories.
 We can parametrize the trajectories using the proper time as follows: 
\begin{align}
	x_r(\tau) = \begin{pmatrix}
		\tau / \sqrt{1-r_i^2/\ell^2} \\ r_i
	\end{pmatrix} \label{eq:trajectory}
\end{align}

These trajectories experience the following proper acceleration, which are different for the different choices of $r_i$:
\begin{align*}
	u^\mu =& \partial_\tau x^\mu = \begin{pmatrix}
		1 / \sqrt{1-r_i^2/\ell^2} \\ 0
	\end{pmatrix}\\
	a^\mu =& \partial_\tau u^\mu + \christ{\mu}{\lambda\nu} u^\lambda u^\nu\\
	a^t =& 2\christ{t}{tr} u^t u^r = 0\\
	a^r =& \christ{r}{tt} u^t u^t + \christ{r}{rr} u^r u^r \\=& \frac{r_i}{\ell^4}(r_i^2-\ell^2) \frac{1}{1-r_i^2/\ell^2} = -\frac{r_i}{\ell^2}\\
	|a|^2(r) =& a_\mu a^\mu=  g_{\mu\nu} a^\mu a^\nu = g_{rr} a^r a^r = \frac{1}{\ell^2-r^2} \frac{r^2}{\ell^2} 
\end{align*}
where the non-vanishing Christoffel symbols in static coordinates are $\christ{t}{tr}=\christ{t}{rt}=\frac{r}{r^2-\ell^2}$, $\christ{r}{tt}=\frac{r}{\ell^4}(r^2-\ell^2)$ and $\christ{r}{rr} =-\frac{r}{r^2-\ell^2}$. For the absolute value of the proper acceleration, we are interested in two special cases:
\begin{align}
    |a|^2(r)&\xrightarrow{r\rightarrow\pm\ell} \infty\\ \quad 
    |a|^2(0)&=0
\end{align}

We can look at the proper acceleration of the same set of trajectories from the embedding space (where all Christoffel symbols vanish identically):

\begin{align}
	x_i^\mu(\tau)=&\begin{pmatrix}
		\sqrt{\ell^2-r_i^2} \sinh(\tau / \sqrt{\ell^2-r_i^2})\\
		 \sqrt{\ell^2-r_i^2} \cosh(\tau / \sqrt{\ell^2-r_i^2})\\
		 r_i 
	\end{pmatrix}\\
	u_i^\mu(\tau) =& \begin{pmatrix} \cosh(\tau / \sqrt{\ell^2-r_i^2}) \\ \sinh(\tau / \sqrt{\ell^2-r_i^2}) \\ 0
	\end{pmatrix}\\
	a_i^\mu(\tau) =& \frac{1}{\sqrt{\ell^2-r_i^2}}\begin{pmatrix}
		 \sinh(\tau / \sqrt{\ell^2-r_i^2})\\
		 \cosh(\tau / \sqrt{\ell^2-r_i^2})\\
		 0
	\end{pmatrix}\\
	a_\mu a^\mu =& \eta_{\mu\nu} a_i^\mu a_i^\nu
	= \frac{1}{\ell^2-r_i^2}
\end{align}

\onecolumngrid
\section{Express the multi-rail state with static patch operators} \label{app:multi-rail-expressions}

Here, we derive an equivalent expression of the initial state of the universe where we use the commutation relation: 
\begin{align}
\left[\hat{a},e^{\lambda \hat{a}^\dagger\hat{b}^\dagger}\right] = \lambda\hat{b}^\dagger  \, e^{ \lambda \hat{a}^\dagger\hat{b}^\dagger}    
\end{align}
The vacuum state takes the following form: 
\begin{align}
    \ket\Omega =& \mathcal{N} \exp\left[\sum_k \int_0^\infty d\omega e^{-\pi\omega}
    (\hat{a}_{k \omega}^{\I})^\dagger (\hat{a}_{k \omega}^{\II})^\dagger \right] \ket{\Omega_{\I}} \otimes \ket{\Omega_{\II}}\\
    =& \mathcal{N}  \prod_{\omega=0}^{\infty} 
    \exp\left[\sum_{k}{e^{-\pi\omega} (\hat{a}_{k \omega}^{\I})^\dagger (\hat{a}_{k \omega}^{\II})^\dagger} \right]\ket{\Omega_{\I}} \otimes \ket{\Omega_{\II}}
\end{align}
We can express the multi-rail state as follows:
\begin{align*}
    \ket\psi =& \sum_{j=1}^{d} c_j (\hat{a}_{k_j \omega_j})^\dagger \ket{\Omega}\\
    =&\sum_{j=1}^{d} c_j \frac{1}{\sqrt{1-e^{-2\pi\omega_i}}} \left[(\hat{a}_{k_j \omega_j}^{\I})^\dagger - e^{-\pi\omega_i} \hat{a}_{k_j \omega_j}^{\II} \right]\ket{\Omega}\\
    =&\sum_{j=1}^{d} c_j \frac{1}{\sqrt{1-e^{-2\pi\omega_i}}} \left[(\hat{a}_{k_j \omega_j}^{\I})^\dagger - e^{-\pi\omega_i} \hat{a}_{k_j \omega_j}^{\II} \right]\mathcal{N}\prod_{\omega=0}^{\infty}  e^{\sum_k e^{-\pi\omega} (\hat{a}_{k \omega}^{\I})^\dagger (\hat{a}_{k \omega}^{\II})^\dagger} \ket{\Omega_{\I}} \otimes \ket{\Omega_{\II}}\\
    =& \sum_{j=1}^{d} \frac{c_j (\hat{a}_{k_j \omega_j}^{\I})^\dagger}{\sqrt{1-e^{-2\pi\omega_j}}} \mathcal{N}\prod_{\omega=0}^{\infty}  e^{\sum_k e^{-\pi\omega} (\hat{a}_{k \omega}^{\I})^\dagger (\hat{a}_{k \omega}^{\II})^\dagger} \ket{\Omega_{\I}} \otimes \ket{\Omega_{\II}} \\
    & -\sum_{j=1}^{d} \frac{c_j  e^{-\pi\omega_j}}{\sqrt{1-e^{-2\pi\omega_j}}} \mathcal{N}\prod_{\omega=0}^{\infty}   \left\{ e^{\sum_k e^{-\pi\omega} (\hat{a}_{k \omega}^{\I})^\dagger (\hat{a}_{k \omega}^{\II})^\dagger} \hat{a}_{k_j \omega_j}^{\II} + \left[\hat{a}_{k_j \omega_j}^{\II}, e^{e^{-\pi\omega} (\hat{a}_{k \omega}^{\I})^\dagger (\hat{a}_{k \omega}^{\II})^\dagger}\right] \right\} \ket{\Omega_{\I}} \otimes \ket{\Omega_{\II}} \\
    =&\sum_{j=1}^{d} \frac{c_j (\hat{a}_{k_j \omega_j}^{\I})^\dagger}{\sqrt{1-e^{-2\pi\omega_j}}} \mathcal{N}\prod_{\omega=0}^{\infty}  e^{\sum_k e^{-\pi\omega} (\hat{a}_{k \omega}^{\I})^\dagger (\hat{a}_{k \omega}^{\II})^\dagger} \ket{\Omega_{\I}} \otimes \ket{\Omega_{\II}} \\
    & -\sum_{j=1}^{d} \frac{c_j  e^{-\pi\omega_j} }{\sqrt{1-e^{-2\pi\omega_j}}}\mathcal{N}  \prod_{\omega=0}^{\infty}  \left[(\hat{a}_{k_j \omega_j}^{\I})^\dagger e^{-\pi\omega_j} e^{\sum_k e^{-\pi\omega} (\hat{a}_{k \omega}^{\I})^\dagger (\hat{a}_{k \omega}^{\II})^\dagger}\right]  \ket{\Omega_{\I}} \otimes \ket{\Omega_{\II}}\\
    =& \sum_{j=1}^{d} \frac{c_j (\hat{a}_{k_j \omega_j}^{\I})^\dagger \left(1-e^{-2\pi\omega_j} \right)}{\sqrt{1-e^{-2\pi\omega_j}}} \ket{\Omega}\\ 
    =& \sum_{j=1}^{d} c_j (\hat{a}_{k_j \omega_j}^{\I})^\dagger \sqrt{1-e^{-2\pi\omega_j}} \ket{\Omega}
\end{align*}

\section{Multi-Index notation} \label{app:multinomial} 
To further simplify the expression of the multi-rail state, we introduce the abbreviation $z=e^{-\pi\omega}$ and expand the exponential function:
				\begin{align}
						\ket{\psi_E} 
						=&\sum_{j=1}^{d} c_j {\sqrt{1-e^{-2\pi{\omega_j}}}}(\hat{a}_{k_j \omega_j}^{\I})^\dagger \mathcal{N}{\prod_{\omega}} \,{\exp}\left[\sum_k e^{-\pi{\omega}}  (\hat{a}_{{k \omega}}^{\I})^\dagger (\hat{a}_{{k \omega}}^{\II})^\dagger \right]  \ket{\Omega_{\I}} \otimes \ket{\Omega_{\II}}\nonumber\\
						=&\sum_{j=1}^{d} c_j {\sqrt{1-z_j^2}}(\hat{a}_{k_j \omega_j}^{\I})^\dagger \mathcal{N} \,\sum_{n=0}^{\infty }\frac{1}{n!}\left({\sum_{\omega}}z  (\hat{a}_{{k\omega}}^{\I})^\dagger (\hat{a}_{{k\omega}}^{\II})^\dagger \right)^{{n}}  \ket{\Omega_{\I}} \otimes \ket{\Omega_{\II}}\nonumber\\
						=& \sum_{j=1}^{d} c_j \sqrt{1-z_j^2}  \sum_{n=0}^\infty  \sum_{L_n} \mathcal{N}\left(\prod_{j=1}^{d}  z^{l_j}\right)  \underset{=\sqrt{l_j+1}\ket{L^{(j)}}_{\I}}{\underbrace{(\hat{a}_{k_j\omega_j}^{\I})^\dagger \ket{l_1 l_2 \dots l_d}_{\I}}} \otimes \underset{=\ket{L}_{\II}}{\underbrace{\ket{l_1 l_2 \dots l_d}_{\II}}} \nonumber\\
						=& \sum_{j=1}^{d} c_j \sqrt{1-z_j^2}  \sum_{n=0}^\infty   z^n \mathcal{N} \sum_{L_n} \sqrt{l_j+1} \ket{L^{(j)}}_{\I} \ket{L}_{\II}\label{eq:altermateEucMultiRail}
					\end{align}
We used the following relations: 
\begin{align*}
	\ket{l_i} =& \frac{1}{\sqrt{l_i!}}(\hat{a}_{k_i\omega_i}^\dagger)^{l_i} \ket{\Omega}\\
	\ket{L} =& \ket{l_1 l_2 \dots l_d}  \\
	\ket{L^{(i)}}_{\I} =& (l_i+1)^{-1/2} (\hat{a}_{k_i\omega_i}^{\I})^\dagger \ket{l_1 l_2 \dots l_d}_{\I} \\
	\sum_{L_n}=&\sum_{l_1+\cdots l_d=n}
\end{align*}
					
					After expanding the exponential function, we applied the multinomial theorem:
					\begin{align}
						\frac{1}{k!}\left( \sum_{i=1}^d \hat{a}_i^\dagger \hat{b}_i^\dagger\right)^k = \sum_{l_1+l_2+\dots+l_d=k} \frac{1}{l_1! l_2! \dots l_d!} (\hat{a_1}^\dagger \hat{b}_1^\dagger)^{l_1}\cdots  (\hat{a_d}^\dagger \hat{b}_d^\dagger)^{l_d}
					\end{align}

\section{Unruh cloning channel} \label{app:cloning-channel}

We take the initial state from eq. (\ref{eq:multi-rail-E-new}) where we consider all $z$ to be equal. The resulting state is:
\begin{align}
	\ket{\psi} 
	= \sum_{j=1}^{d} c_j (1-z^2)^{(d+1)/2}  \sum_{n=0}^\infty z^n   \sum_{L_n} \sqrt{l_j+1} \ket{L^{(j)}}_{\I} \ket{L}_{\II}
\end{align}
With this expression of $\ket\psi$ we derive the Unruh channel from \ref{eq:cloning-channel} $$	\mathcal{E}(\psi)  = \tr_{\II}( \ket{\psi}\bra\psi )= \rho_{\I} .$$ In doing so, we split the channel into two parts: the diagonal part, where identical frequencies are excited, and the off-diagonal part, where different frequencies are excited. This distinction affects the index we call $k$, which is the mode excited in the multi-rail state. 
\begin{align}
	\rho_{\I} 
	=& \tr_{\II} \left[ \left\{\sum_{j=1}^{d} c_j (1-z^2)^{(d+1)/2}  \sum_{n=0}^\infty z^n   \sum_{L_n} \sqrt{l_j+1} \ket{L^{(j)}}_{\I} \ket{L}_{\II}\right\} \right.\\&\left. \cdot \, \left\{\sum_{\tilde{j}=1}^{d} c_{\tilde{j}}^\ast (1-z^2)^{(d+1)/2}  \sum_{n=0}^\infty z^n   \sum_{L_n} \sqrt{l_{\tilde{j}}+1} \bra{L^{(\tilde{j})}}_{\I} \bra{L}_{\II}\right\} \right]\\
	=& (1-z^2)^{d+1} \sum_{n=0}^{\infty} z^{2n} \sigma_{\I}^{(n)} 
\end{align}
with \begin{align}
    \sigma_{\I}^{(n)} =& \sum_{j=1}^d |c_j|^2(l_j+1)\sum_{L_n} \ket{L^{(j)}}\bra{L^{(j)}}_{\I} + \sum_{\substack{j,\tilde{j}=1\\k\neq\tilde{j}}}^d c_j c_{\tilde{j}}\sqrt{l_j+1}\sqrt{l_{\tilde{j}}+1}\sum_{L_n} \ket{L^{(j)}}\bra{L^{(\tilde j)}}_{\I}
\end{align}

\twocolumngrid
	\end{appendix}

	\twocolumngrid
	\bibliography{bibliography}
\end{document}